\documentclass[useAMS,usenatbib]{mn2e}
\pdfoutput=1
\usepackage{color}
\usepackage{times}
\usepackage{amsmath}
\usepackage{amssymb}
\usepackage{amsbsy}
\usepackage{graphicx}
\usepackage{subfigure}
\usepackage{paralist}

\def\kpc{\,{\rm kpc}}
\def\mas{\,{\rm mas}}
\def\pc{\,{\rm pc}}

\def\masyr{\,{\rm mas\,yr^{-1}}}
\def\kms{\,{\rm km\,s^{-1}}}
\def\rad{\,\rm rad}
\def\Gyr{\,{\rm Gyr}}

\def\percent{\text{ per cent}}

\title[Constraining the Galactic potential with streams]{Stream-orbit misalignment II: A new algorithm to constrain the Galactic potential}
\author[J. L. Sanders and J. J. Binney]{Jason L. Sanders$^1$\thanks{E-mail: jason.sanders@physics.ox.ac.uk} and James Binney$^1$\\
$^1$ Rudolf Peierls Centre for Theoretical Physics, Keble Road, Oxford OX1
3NP, UK\\ 
}

\pagerange{\pageref{firstpage}--\pageref{lastpage}} \pubyear{2012}
\label{firstpage}

\begin{document}
\maketitle

\begin{abstract}
In the first of these two papers we demonstrated that assuming streams
delineate orbits can lead to order one errors in potential parameters for
realistic Galactic potentials. Motivated by the need for an improvement on
orbit-fitting, we now present an algorithm for constraining the Galactic
potential using tidal streams without assuming that streams delineate orbits.
This approach is independent of the progenitor mass so is valid for all
observed tidal streams. The method makes heavy use of angle-action variables
and seeks the potential which recovers the expected correlations in angle
space. We demonstrate that the method can correctly recover the parameters of
a simple two-parameter logarithmic potential by analysing an N-body
simulation of a stream. We investigate the magnitude of the errors in
observational data for which the method can still recover the correct
potential and compare this to current and future errors in data. The errors
in the observables of individual stars for current and near future data are
shown to be too large for the direct use of this method, but when the data
are  averaged in bins on the sky, the resulting averaged data are accurate
enough to constrain correctly the potential parameters for achievable
observational errors. From pseudo-data with errors comparable to those that
will be furnished in the era of {\it Gaia} ($20\percent$ distance errors,
$1.2\masyr$ proper motion errors, and $10\kms$ line-of-sight velocity errors)
we recover the circular velocity, $V_c=220\kms$, and the flattening of the
potential, $q=0.9$,
to be $V_c=223\pm10\kms$ and $q=0.91\pm0.09$.
\end{abstract}


\section{Introduction}

The study of the Galactic halo is crucial to the understanding of the dark-matter content of the Galaxy. The dynamics of halo stars reflect the shape of the Galactic potential on large scales and hence the underlying dark-matter distribution. Recent large photometric surveys have revealed many stream-like structures in the halo \citep{Belokurov2006-FoS}, which should prove useful in this respect. These tidal streams are believed to be formed by stars being stripped from their progenitor by the tidal forces of the Galactic potential. The formation of the stream reflects the underlying Galactic potential, such that the resulting structure may be used to constrain the potential. 

One way of using streams to constrain the Galactic potential is to assume that a stream delineates an orbit. This assumption has been used by many authors in the development of stream-fitting algorithms \citep{Jin2007,Binney2008,EyreBinney2009B,EyreBinney2009A} and their application to real data \citep{Willett2009,Koposov2010}. However, not until recently was the assumption that a stream delineates an orbit properly questioned by \citet{EyreBinney2011}. These authors demonstrated that a misalignment between the progenitor orbit and stream can lead to significant errors in the parameters of simple potentials. \citet[hereafter Paper I]{SandersBinney2013} expanded on this by discussing the misalignment in realistic Galactic potentials for real streams, and demonstrated that orbit-fitting is not valid for many known streams. In particular order-one errors in the halo parameters of a realistic Galactic potential can arise if orbit-fitting is applied to known Galactic streams. Clearly, an improvement over orbit-fitting is required, which accounts for the stream-orbit misalignment.

\citet{Johnston1999} accounted for the misalignment between the stream and progenitor orbit by calculating a progenitor-mass-dependent energy offset at pericentre. The observed stream stars are assigned an energy in this range, and the stream is integrated backwards for a Galactic lifetime. The quality of the trial potential is assessed by the number of `captured' particles at time $t=0$. \citet{Varghese2011} developed a similar method which took into account the misalignment between the stream and progenitor orbit by correcting a proposed orbit track in real-space with a progenitor-mass dependent term. To fully model stream formation with limited assumptions we must turn to N-body models. \citet{Law2005}, \citet{Fellhauer2006} and \citet{LawMajewski2010} have all employed N-body models of the tidal disruption of the Sagittarius dwarf to produce constraints on the Galactic potential. However, full searches over the entire parameter space, both potential parameters and initial cluster conditions, are very expensive as we must create a new simulation each time, and it is difficult to assess how well a given simulation fits the data. \citet{Penarrubia2012} present an entropy-based algorithm for constraining the Galactic potential with tidal streams. Assuming the distribution function is separable in energy and position, the best-fit potential is the one which minimises the entropy of the energy distribution, or, equivalently, minimises the range of energies of the stream members. These are all valid alternatives to orbit-fitting.

Angle-action coordinates are very useful quantities in galactic dynamics, and
with more and more methods for their calculation becoming available
\citep{McMillanBinney2008,Sanders2012,Binney2012} their use is becoming a
practical possibility. In particular, stream formation is very simply
expressed in the angle-action formalism \bibpunct[;
]{(}{)}{;}{a}{}{;}\citep[Paper I]{Tremaine1999,HelmiWhite1999,EyreBinney2011}\bibpunct[;
]{(}{)}{;}{a}{}{,}. The structure of a tidal stream in angle-action space
will only be physically correct if we have used the correct potential,
regardless of whether the stream delineates an orbit or not. This formalism
provides us with a route to develop a new algorithm to constrain the Galactic
potential: we search for the potential which produces the correct
correlations in angle-action space. In Section~\ref{Recap} we recap the
angle-action framework of stream formation, and more closely inspect an
N-body simulation to motivate and understand this formalism. In
Section~\ref{Algorithm} we discuss how we can improve on orbit-fitting by
exploiting the known angle-action correlations for streams and arrive at a
suitable algorithm to constrain the potential. In Section~\ref{Test} we test
the algorithm by attempting to recover the potential from the simulation.

Due to their large distances from the Sun, observations of tidal streams have large errors. It is therefore important that any stream-fitting algorithm can cope with appropriate observational errors. Some authors have tested their methods on simulated observational data with appropriate errors \citep{Binney2008,EyreBinney2009B}. In Section~\ref{Errors}, we conclude with a full discussion of the effect of observational errors on our proposed algorithm, and compare to the errors from current and future surveys.

\section{Streams in angle-action space}\label{Recap}
In Paper I we presented the angle-action formalism of stream formation. In this formalism streams form due to stars being on different orbits. Importantly this can lead to a misalignment, $\varphi$, between the stream and the underlying progenitor orbit given by
\begin{equation}
\varphi \equiv \arccos\Big(\hat{\boldsymbol{\Omega}}_0\cdot\hat{\boldsymbol{e}}_1\Big),
\label{Misalignment}
\end{equation}
where $\boldsymbol{\Omega}_0$ is the frequency vector of the progenitor orbit. $\hat{\boldsymbol{e}}_1$ is the principal eigenvector of the Hessian, $\boldsymbol{D}$, given by
\begin{equation}
D_{ij}(\boldsymbol{J})=\frac{\partial^2 H}{\partial J_i \partial J_j}.
\end{equation}
This misalignment only depends upon the progenitor orbit, and hence the potential, and not the progenitor mass. In Paper I we demonstrated that a non-zero misalignment angle leads to biases in the estimation of potential parameters when fitting orbits to streams. We therefore require an alternative to orbit-fitting, and the angle-action formalism provides us with a clear route.

%In deriving equation~\eqref{Misalignment} we approximated the frequency of a stream star as a first order Taylor expansion about the progenitor frequency. 

For long narrow streams to form, the Hessian must be dominated by a single eigenvalue, $\lambda_1$, with corresponding eigenvector, $\hat{\boldsymbol{e}}_1$. Under this approximation both the angle and frequency differences of all the stars in the stream lie along the same straight line:
\begin{equation}
\frac{\Delta\boldsymbol{\theta}}{t}\approx\Delta\boldsymbol{\Omega}\approx \hat{\boldsymbol{e}}_1 (\lambda_1 \hat{\boldsymbol{e}}_1\cdot\Delta\boldsymbol{J}).
\label{AngFreqAlignment}
\end{equation}
In the correct potential this correlation between the angle and frequency-space structure of the stream should be apparent. Inspecting the stream in angle-action space for a trial potential should tell us whether this potential is the true potential.

In practice, the actions, angles and frequencies are difficult to calculate for a general potential. \cite{Sanders2012} presents a method for finding actions and angles in a general axisymmetric potential by locally fitting a St\"ackel potential. The St\"ackel-fitting approach can be simply extended to estimate the frequencies of an orbit. We present this extension in Appendix~\ref{StackelFit} and there also discuss refinements to the algorithm which are appropriate for the problem in hand.

\subsection{A simulation}
In Paper I we introduced a range of N-body simulations to show that the angle-action formalism is mass-independent for all interesting progenitor mass scales. Before discussing the details of the proposed algorithm we briefly introduce one of these N-body stream simulations to motivate the discussion. We take the lowest mass stream from Paper I which was produced as follows: we construct a stream by evolving an N-body simulation of a King cluster on a stream-like orbit in the logarithmic potential using the code {\textsc{gyrfalcON}} \citep{Dehnen2000,Dehnen2002}. The simple two-parameter logarithmic potential is given by
\begin{equation}
\Phi(R,z) = \frac{V_c^2}{2}\ln\Big(R^2+\frac{z^2}{q^2}\Big),
\label{LogPot}
\end{equation}
where $V_c$ is the asymptotic circular speed and $q$ is the flattening parameter. We choose $V_c=220\kms$ and $q=0.9$ which gives a good representation of the potential of the Milky Way \citep{Koposov2010}. King models \citep{King1966} are characterised by three free parameters: the ratio of central potential to squared-velocity parameter, $W_0 = \Psi_0/\sigma^2$, the cluster mass, $M_c$, and a tidal limiting radius, $r_t$. We set $W_0=2.0$ and $M_c=2\times10^4M_\odot$ and seed the cluster with $N=10000$ particles. Following \citet{Dehnen2004} we relate $r_t$ to the pericentric radius of the orbit, $r_p$, via
\begin{equation}
r_t ^3\approx\frac{GM_c}{V_c^2}r_p^2.
\end{equation}

\begin{figure}
$$\includegraphics{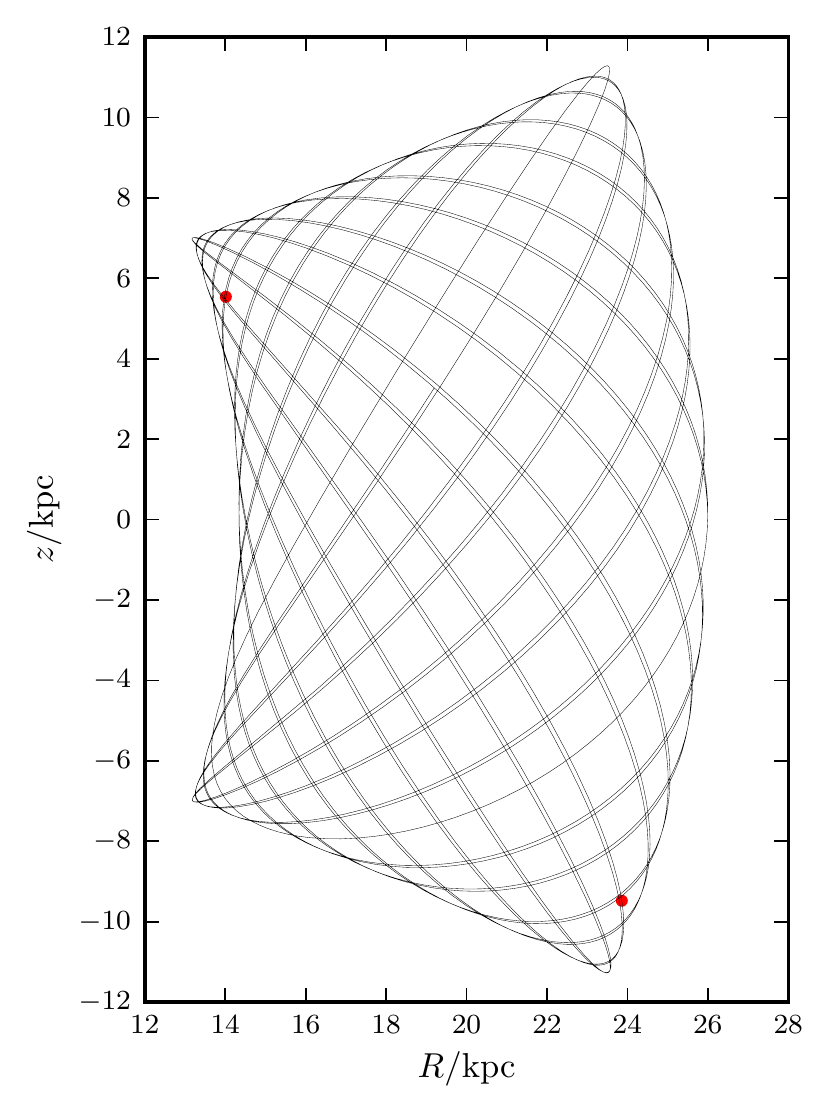}$$
\caption{Progenitor orbit used for the simulation in Section~\ref{Test}. The positions of the cluster at $t=4.02\Gyr$ and $t=4.27\Gyr$ are marked by red dots.}
\label{Orbit}

\end{figure}

We place the cluster at the apocentre of the orbit shown in Fig.~\ref{Orbit}, which has initial conditions $(R,z)=(26.0,0.0)\kpc$ and $(U,V,W)=(0.0,141.8,83.1)\kms$. Positive $U$ is towards the Galactic centre and positive $V$ is in the direction of the Galactic rotation at the Sun. This orbit has $r_p \approx 14\kpc$, and was chosen due to its similarity to the GD-1 orbit found by \citet{Koposov2010}. We evolve the simulation for $t=5\Gyr$ (approximately $12$ radial oscillations of the progenitor)  The chosen cluster model parameters along with quantities derived from these parameters are given in Table~\ref{KingModelTable}.

\begin{figure}
$$\includegraphics{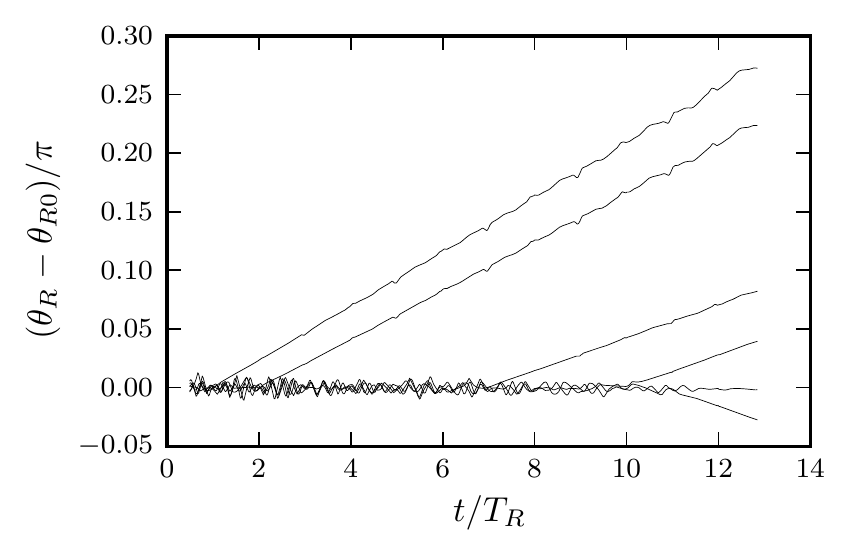}$$
\caption{Difference between the radial angle of the progenitor and six particles selected from the simulation. At early times the particles oscillate inside the cluster until they are released at pericentre (given by the units on the $x$-axis). Particles are stripped symmetrically in angle space. After release the particles evolve freely in the external Galactic potential. The small blips in the plot are due to numerical errors introduced by the St\"ackel-fitting algorithm. The particles which are released first have larger frequency differences (i.e. steeper slopes in the plot) than those released at later times.}
\label{AngleswithTimes}
\end{figure}

Fig.~\ref{AngleswithTimes} shows the time evolution of the angle differences between the progenitor and several particles in the simulation. Each particle oscillates in the cluster until it escapes at pericentre, after which it moves as a free particle in the Galactic potential. The slope of the particle's motion in angle-space is given by the frequency. We see that the first particles to leave have a higher frequency than those released at later times. This is because the first particles to leave require a higher energy to escape the cluster. One unexpected feature of Fig.~\ref{AngleswithTimes} is the noise in the angle calculation when the particles are moving freely. There are small bumps in the angle difference at each orbit pericentre, which alternate in sign and increase in magnitude with time. This is due to errors introduced by the St\"ackel-fitting algorithm for estimating the angles and frequencies, which we discuss later.

\begin{figure}
\begin{minipage}{0.5\columnwidth}
$$\includegraphics{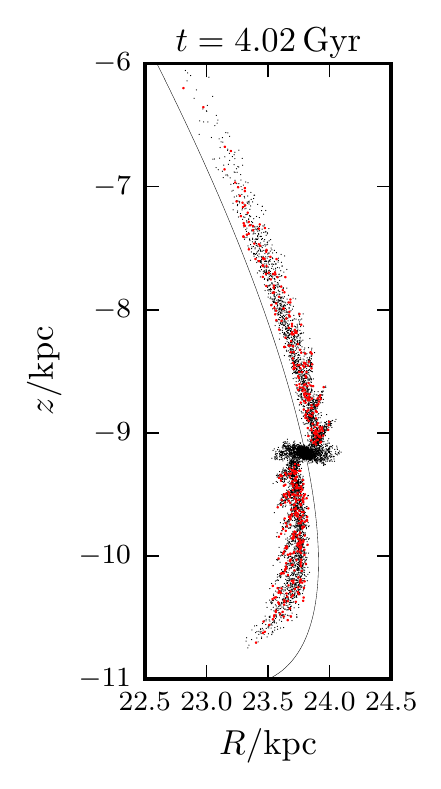}$$
\end{minipage}
\begin{minipage}{0.5\columnwidth}
$$\includegraphics{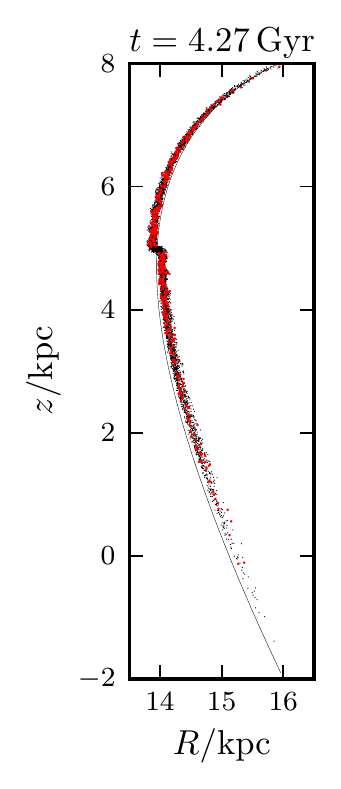}$$
\end{minipage}
\caption{Cluster at $t=4.02\Gyr$ (just after 10th apocentric passage) and $t=4.27\Gyr$ (just after subsequent pericentric passage). The solid line is the orbit of the progenitor and the red dots are the sample of $500$ stars used to constrain the potential in Section~\ref{Test}.}
\label{ClusterEndState}
\end{figure}

\begin{figure}
$$\includegraphics{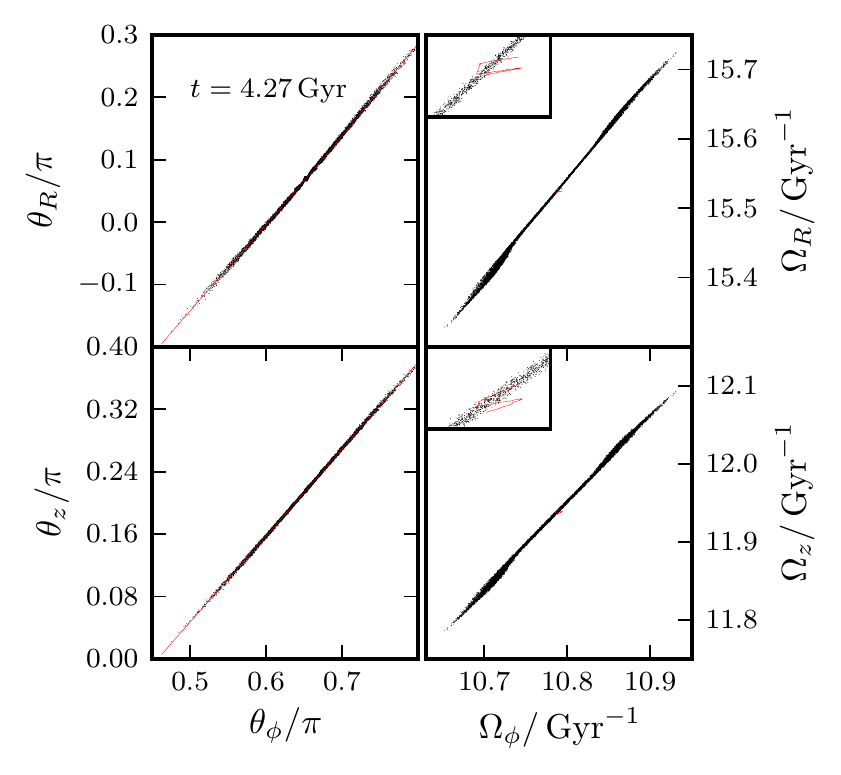}$$
\caption{Cluster at $t=4.27\Gyr$ (just after the 11th pericentre passage) in angle and frequency space. The red line is the projection of the progenitor orbit shown in Fig.~\ref{ClusterEndState} into angle space and frequency space. The angles and frequencies were calculated using the true potential. The inset plots show a zoom-in of the frequency space so the orbit projections are more easily visible. In frequency-space, errors in the determination of frequencies cause the red points to form a line rather than all coincide as they should.}
\label{ClusterEndStateAA}
\end{figure}

In the right panel of Fig.~\ref{ClusterEndState} we show the formed stream in real-space at $t=4.27\Gyr$ (just after the $11$th pericentric passage). Fig.~\ref{ClusterEndStateAA} shows the corresponding angle-space and frequency-space structures calculated using the correct potential. We see that as predicted by equation~\eqref{AngFreqAlignment} the stream stars all lie along a straight line in both angle and frequency-space. In Fig.~\ref{ClusterEndStateAA} we see that the spread in frequencies in the cluster is much larger than the spread in the estimated frequencies along the progenitor orbit (given by the red line). This suggests that the St\"ackel-fitting algorithm is sufficiently accurate. However, the broadening at the extremes of frequency distributions indicates that there is a systematic error present which will skew the gradient estimation. A fuller discussion of the errors is given later.


\begin{table}
\caption{Parameters of King Model used in the simulation detailed in Section~\ref{Test}. $\epsilon$ is the softening parameter.}
\begin{tabular}{l|llllll}
$N$&$W_0$&$\frac{M_c}{M_\odot}$&$r_p/\kpc$&$r_t/\kpc$&$\sigma/\kms$&$\epsilon/\pc$\\
\hline\\
10000&2.0&$2\times10^4$&$14$&$0.07$&$1.39$&$1.5$\\
\hline
\end{tabular}
\label{KingModelTable}
\end{table}

\section{Algorithm}\label{Algorithm}
The N-body stream observations now provide us with a way to utilise the angle-action formalism to constrain the potential. Equation~\eqref{AngFreqAlignment} states that the frequency and angle differences of all the stars in the stream  must lie along $\hat{\boldsymbol{e}}_1$, and this is observed in the above N-body simulation (Fig.~\ref{ClusterEndStateAA}). For all the stars we can calculate the angles and frequencies and then obtain independent estimates  $\hat{\boldsymbol{e}}^\theta_1$ and $\hat{\boldsymbol{e}}^\Omega_1$ of $\hat{\boldsymbol{e}}_1$ by performing linear fits to the angles and frequencies. The potential closest to the Galaxy's potential is then the potential which maximises $\hat{\boldsymbol{e}}^\theta_1\cdot\hat{\boldsymbol{e}}^\Omega_1$. The present kinematics are given by $\hat{\boldsymbol{e}}^\Omega_1$ -- it gives us the direction that stars will move in if the true potential suddenly changed to the trial potential. However, $\hat{\boldsymbol{e}}^\theta_1$ gives a measure of the positions the stars have reached by moving in the true potential i.e. the history of the stars. In the true potential the current motions of stars (along $\hat{\boldsymbol{e}}^\Omega_1$) will match the positions they have in fact reached (along $\hat{\boldsymbol{e}}^\theta_1$).

Equation~\eqref{AngFreqAlignment} was derived on the assumption that the stream structure is isotropic in action-space. In Paper I we saw, in general, there are anisotropies in the action-space distribution, which cause a deviation of the frequency structure from the principal eigenvector of the Hessian. For long thin streams the Hessian must have a large principal-to-second-eigenvalue ratio. Therefore, we expect that whatever the shape of the action-space distribution, the resulting frequency-distribution will be highly elongated, although not necessarily along the principal eigenvector of the Hessian. The approach suggested here is comparing the angle and frequency distributions which should have the same principal axis in the correct potential. Therefore, the approach should be insensitive to anisotropies in action-space if we are observing long thin streams.

This approach has several clear advantages:
\begin{inparaenum}
\item We haven't assumed that the stars in the stream delineate an orbit.
\item The position of the progenitor is irrelevant. It reduces to a constant in the linear fits.
\item We can use any subsection of the stream.
\item The time since each star was stripped does not matter. Changing the assumed stripping time only moves stars up and down the straight line in angle space and so does not affect the linear fit.
\item Only a single calculation is required for each trial potential, whereas when orbits are fitted in real space, orbits must be generated in each potential until the best fit is found.
\item The quality of the potential is determined solely by the degree of misalignment, so it is an unambiguous single measure of the fit. When fitting orbits, the quality of the potential is evaluated by measuring the proximity of the orbit found to the data in the 6D phase space, so an arbitrary metric must be introduced to relate position and velocity differences.
\end{inparaenum}

However, our approach suffers from the disadvantages that 
\begin{inparaenum}
\item the angles and frequencies can only be found with full six-dimensional phase-space information,
\item stream data are often poor and it is not clear how the errors affect the accuracy of the method,
\item and unlike orbit fitting, we must process the data before assessing the quality of the potential, as opposed to directly checking the quality of the fits in real space.
\end{inparaenum}

In what follows we will assume that we have full phase-space information for all the stars in the stream, so we are not concerned with the first of these points. In Section~\ref{Errors} we explore how errors in the 6D data for the members of the stream affects the accuracy of the method. We summarise the above discussion into the simple algorithm presented in Table~\ref{AlgoTable}.

\newcommand*{\MyIndent}{\hspace*{0.5cm}}
\begin{table}
\caption{Algorithm to find the best-fit potential given stream data.}
\begin{tabular}{ll}
\hline
\multicolumn{2}{l}{Given a set of 6D phase-space points $\boldsymbol{x}_i$ we}\\
\hline
$1.$&pick a trial potential $\Phi_{\rmn{trial}}$,\\
\vspace{0.1cm}
$2.$&\begin{minipage}[t]{.85\columnwidth}
	calculate the angles $\boldsymbol{\theta}_i(\boldsymbol{x}_i,\Phi_{\rmn{trial}})$ and frequencies $\boldsymbol{\Omega}_i(\boldsymbol{x}_i,\Phi_{\rmn{trial}})$ using the local St\"ackel-fitting algorithm,
	\end{minipage}\\
	\vspace{0.1cm}
$3.$&\begin{minipage}[t]{.85\columnwidth}
	fit four straight lines to the graphs of
	\begin{enumerate}
	{\setlength\itemindent{12pt} \item $\Omega_\phi$ against $\Omega_R$, }
	{\setlength\itemindent{12pt} \item $\Omega_\phi$ against $\Omega_z$, }
	{\setlength\itemindent{12pt} \item $\theta_\phi$ against $\theta_R$,}
	{\setlength\itemindent{12pt} \item and $\theta_\phi$ against $\theta_z$}
	\end{enumerate}
	to find the gradients denoted by $a_1, a_2, b_1, b_2$,
	\end{minipage}\\
	\vspace{0.1cm}
$4.$&
\begin{minipage}[t]{.85\columnwidth}
calculate the quantity 
\[\cos\psi \equiv \displaystyle\frac{a_1b_1+a_2b_2+1}{\sqrt{(a_1^2+a_2^2+1)(b_1^2+b_2^2+1)}},\]
\end{minipage}
\\
\vspace{0.1cm}
$5.$&\begin{minipage}[t]{.85\columnwidth}
pick a new trial potential and repeat until the angle, $\psi$, between the vectors is minimised.\end{minipage}\\
\hline
\end{tabular}
\label{AlgoTable}
\end{table}

\section{Test}\label{Test}
We test the algorithm of Table~\ref{AlgoTable} by seeing how accurately the method recovers the chosen parameters of the potential from the N-body simulation. We examine the stream at two times - once at $t=4.02\Gyr$, just after the tenth apocentric passage of the progenitor, and once at $t=4.27\Gyr$, just after the subsequent pericentric passage. From these snapshots we form samples of stream stars by first removing the remnant of the progenitor and then randomly sampling $500$ of the remaining $\sim5000$ stars. The real-space structures of the disrupted cluster and these samples are shown in Fig.~\ref{ClusterEndState}.

\begin{figure}
$$\includegraphics{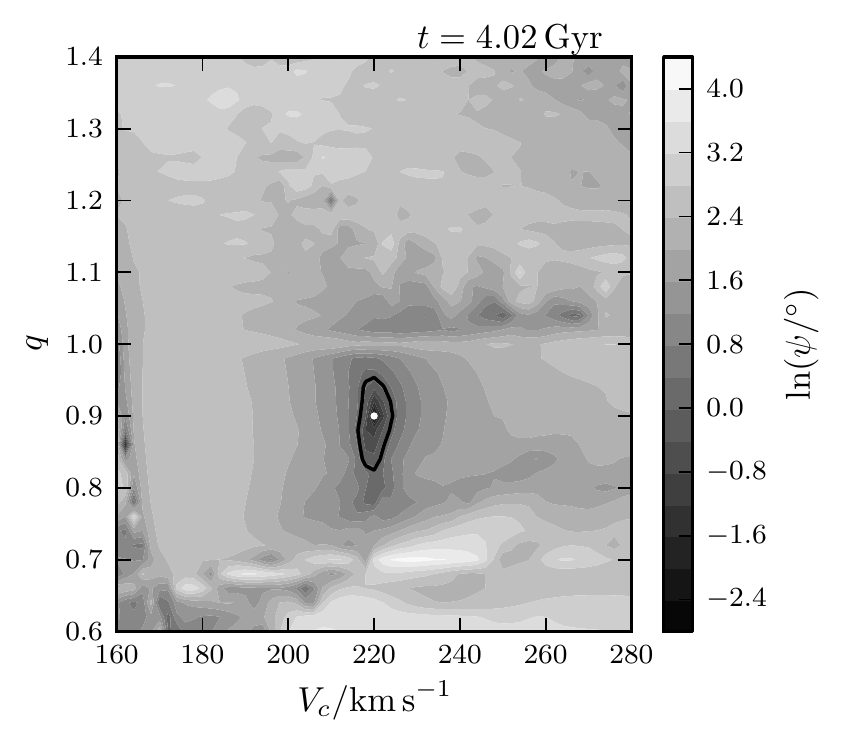}$$
$$\includegraphics{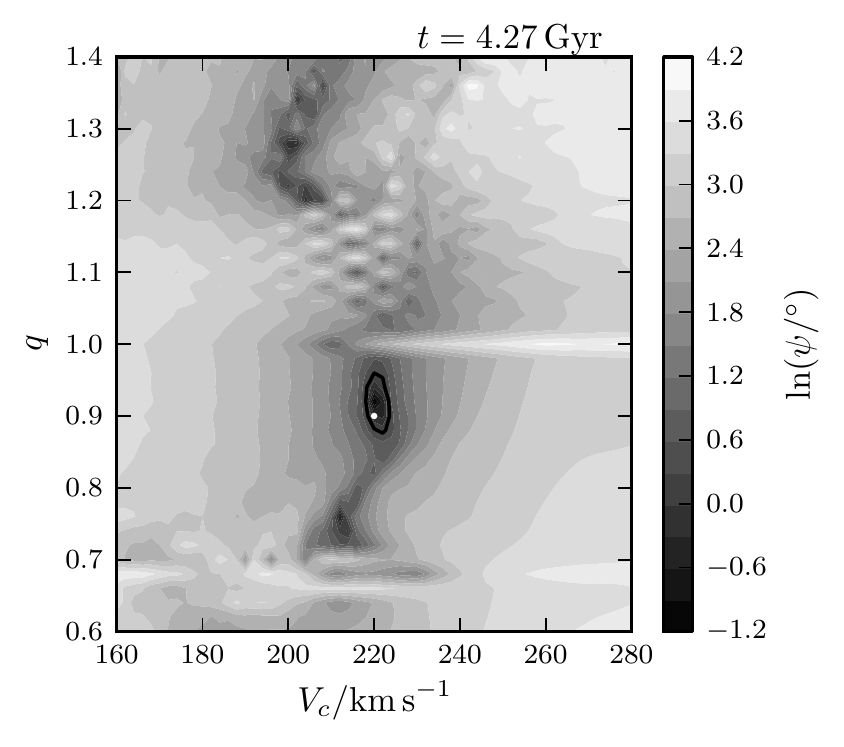}$$
\caption{Misalignment between the angle difference and frequency difference vectors. The results at $t=4.02\Gyr$ (just after the 10th apocentric passage) are shown in the top panel and at $t=4.27\Gyr$ (just after the subsequent pericentric passage) in the bottom panel. The true potential parameters are $V_c=220\kms,q=0.9$ marked by a white dot. The thick black lines mark an error contour showing the uncertainty in the position of the found minimum.}
\label{Results}
\end{figure}

We now use the two samples to diagnose the potential. For each trial pair of $(V_c,q)$ we find the misalignment, $\psi$, between the directions of the angle difference vector and the frequency difference vector. We pick trial pairs of parameters by exploring a regular grid around the true point ($V_c=220\kms, q=0.9$). Fig.~\ref{Results} is a plot of the misalignment in the plane of the parameters.

In the apocentric case (top panel) the global minimum occurs at $V_c=220.3\kms, q=0.899$. The landscape around the minimum is quite complex. There are a series of local minima for $q>1.0$. Therefore, care will be needed when searching for the minimum automatically. In the pericentric case (bottom panel) we do not recover the true potential parameters as successfully: the global minimum lies at $V_c=220.4\kms, q=0.915$ and the $\psi$ landscape is significantly more complex. Again, for $q>1.0$ there are a series of local minima. Worringly, there is another deep local minimum at $V_c\approx 212\kms$, $q\approx0.74$. Clearly there are sources of error in the algorithm which are systematically shifting the best-fit value of $q$ in the pericentric case. Before applying this algorithm to more realistic data sets, we must understand the cause of this systematic shift. More generally, we need a method to assess the magnitude of the systematic error in the algorithm when the underlying potential parameters are not known. In the following section we discuss the causes of the error and present a method to estimate its magnitude.

\subsection{Error estimation}\label{ErrorEstimation}

Even in the correct potential with perfect data the particles will not lie on exact straight lines in angle and frequency space on account of systematic errors introduced by the approximations used including the St\"ackel-fitting algorithm. In the above test we have seen that this can lead to a slight misdiagnosis of the underlying potential parameters. When we include observational errors, the error ellipse obtained by resampling the random errors may not encompass the true parameters due to these systematic errors. We need to quantify the accuracy of the method even when perfect data are used. A naive error may be taken from the error in the linear regression. However, this error only takes account of the random errors and ignores any systematics. In Paper I we discussed the effects of the progenitor mass on the angle-action framework, and hence the algorithm presented in this paper. It was shown that the approach is mass-independent up to about $10^9M_\odot$. Therefore, we expect the self-gravity, the finite cluster size and the linear approximation (see equation~\eqref{AngFreqAlignment}) to not affect the algorithm.

There are still two sources of systematic error in the method. The first is
that the action-space structure is not perfectly isotropic. In Paper I we saw
that particles are not isotropically stripped, but rather escape through the
Lagrange points. This led to the entire stream structure being rotated in the
angle and frequency space. \citet{EyreBinney2011} showed that the
action-space of a tidally-disrupted cluster has a `bow-tie' structure. When
this action-space structure is transformed into frequency space via the
Hessian (see equation~\eqref{AngFreqAlignment}), the broadness of the wings
of the action-space distribution is still apparent at the extremes of the
frequency structure. This is due to the smaller eigenvalues of the Hessian.
This effect is small as we are considering a Hessian with a ratio of
eigenvalues, $\lambda_1/\lambda_2\approx30$. However, this effect should also
be observed in the angle-space structure. As we are comparing the angle and
frequency-space structures, this shouldn't have a significant impact on the
implementation of the algorithm.

% Also, despite the action distribution not being isotropic, for typical progenitor orbits the spread in each action is approximately the same \citep[Paper I]{EyreBinney2011}. This effect should affect both the angle and frequency-space distributions. As our goodness-of-fit measure is a comparison of the two we should not be too sensitive to these effects.

Second, the St\"ackel-fitting algorithm introduces systematic errors. The errors depend upon the orbital phase, such that the errors in the frequency differences between the progenitor and the stream particles exhibit a characteristic beating with time. The particles which were released recently are still approximately in phase with the progenitor, whilst those at the extremes of the frequency distribution, which were released many pericentres ago, have drifted out of phase with the progenitor. Therefore, the St\"ackel-fitting algorithm introduces a larger error in the frequency difference for the particles at the extremes, which leads to a broadening of the distribution at the extremes. In Appendix~\ref{ErrorsAppendix} we give a full discussion.

In conclusion, the extremes of the frequency distribution are broader, and so less reliable, than the central portion of the stream, due to systematic errors introduced by the St\"ackel-fitting algorithm and anisotropies in the action-space distribution. Therefore, we estimate the magnitude of the error in the parameters from the spread in gradients that can be obtained by considering the angle and frequency structure of different parts of the stream in the best-fit potential. Specifically, we take the error in the gradient to be the difference between this gradient and that calculated using the central particles nearest to the progenitor. We may then find a `threshold angle', $\psi_T$, as the angle between the two particle distributions. A threshold angle may be found for both the angle- and frequency-space distributions. However, in the simulation, the frequency distribution is the main cause of error so we only calculate a threshold angle for this distribution. This threshold angle gives an estimate of the minimum angle that we can reliably calculate. The range of valid parameters are those for which $\psi<\psi_T$, which define an error contour in the potential parameter space. Inside the error contour the angle between the frequency and angle distributions is smaller than the width of the frequency distribution.

This also helps to solve an additional problem with the algorithm. As we are blindly fitting straight lines to distributions which may not have a linear structure, there may be some data sets for which the best-fit straight lines fortuitously give a minimum in $\cos\psi$. However, if we calculate an error contour associated with this minimum we expect the error to be very large. Therefore, we should be able to rule out this fortuitous minimum without explicit inspection.

Applying this error estimation method to the data in
Fig.~\ref{ClusterEndStateAA}, we plot an error contour in the parameter
plane, given by the thick black curve in Fig.~\ref{Results}. Assuming the errors in each parameter estimate are independent, we estimate the parameters in the apocentric case as $V_c=(220\pm4)\kms, q=(0.90\pm0.07)$ in good agreement with the truth.
Similarly for the pericentric case,  the parameter estimates are $V_c=(220\pm3)\kms, q=(0.92\pm0.04)$. 

We have shown that we can use our algorithm to constrain the parameters of a
simple potential using an error-free sample of $500$ stars from a stream
simulation. When the stream is observed at either apocentre or pericentre, the error contour
is elongated along the $q$-direction. We can constrain the circular speed of
the potential more accurately than the shape of the potential. The recovery
of the parameters at apocentre is slightly more successful than at
pericentre. However, the errors at apocentre are slightly larger.

\section{Errors in stream data}\label{Errors}
We have assumed in the above test that the input data are perfect. Obviously
this check that the algorithm works in the most optimistic situation is necessary, but it does not give an indication of how the algorithm will perform on a real data set. To give a more realistic test we now add errors to the data and rerun the algorithm. We simulate an observation of the particles at $t=4.27\Gyr$ in the simulation from the position of the Sun $(R,z)=(8.0,0.0)\kpc$, which has velocity $(U,V,W)=(11.1,232.4,7.25)\kms$ \citep{SchonrichBinney2010}. There is considerable uncertainty in the circular speed at the Sun \bibpunct[ ]{(}{)}{;}{a}{}{}\citep[see ][ for a summary]{Bovy2012}\bibpunct[; ]{(}{)}{;}{a}{}{;}, but for a fixed solar position, the solar $V$ is well constrained by the motion of Sgr A* \citep{Reid2004}. Therefore, in what follows, we fix the velocity of the Sun, irrespective of the choice of potential, as this more accurately simulates a realistic application of the above algorithm.

The pericentric snapshot is chosen as the GD-1 stream is currently around pericentre. At pericentre the stellar density in the stream is increased, so the overdensity is more likely to be observed. However, this effect is counteracted by the short time a stream spends around pericentre. We project the positions and velocities of the particles relative to the Sun into observable space on the sky -- sky position $(l,b)$, distance $s$, proper motions $(\mu_l,\mu_b)$ and line-of-sight velocity $v_{||}$. The particles are then scattered in observable space by appropriate Gaussian errors $(\sigma_{l,b},\sigma_s,\sigma_{\mu},\sigma_{||})$ to form an `observed' data set.

The most accurate data are obtained from streams that are closest to the Sun. For instance GD-1 lies between 8 and 12$\kpc$ away from the Sun. Therefore when we perform the above data scattering we place the Sun as close to the centre of the stream as possible. This involves rotating the stream around the Galaxy until the centre of the stream lies at the same azimuthal angle as the Sun. This better simulates observations which could feasibly be performed.

\begin{figure}
$$\includegraphics{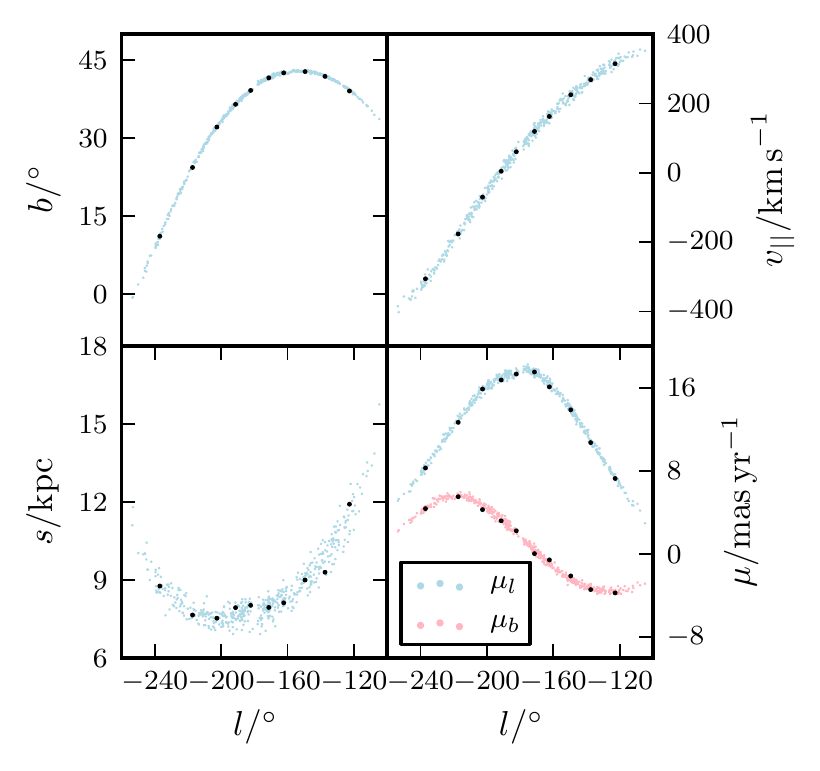}$$
\caption{Stream sample at $t=4.27\Gyr$ (just after 11th pericentric passage). The small blue and red points show the sample observed from the Sun with observational errors O2 given in Table~\ref{ObsErrors} and the larger black points show the reduced data set produced by averaging the data in observable space on the sky with errors OA3.}
\label{Observed}
\end{figure}

We make `observations' with the errors listed in Table~\ref{ObsErrors}, along with errors in current and future data. The data from the O2 observation are shown in Fig.~\ref{Observed}. We now use each of these data sets to constrain the potential. We transform the data set back into the Galactocentric coordinate system, and determine the angle and frequency distributions and $\psi$ for each trial potential. We use the Nelder Mead algorithm to find the maximum of $\cos\psi$ as a function of $q$ and $V_c$. We restrict the range of parameter space explored to $0.7<q<1.1$ and $170\kms<V_c<270\kms$. The initial point passed to the algorithm is the true minimum position. We estimate the error in the position of the minimum for each set of observational errors by repeating this process $100$ times and finding the minimum of each data set. These points sample the distribution of the parameters given the observational errors. Using these points we can reconstruct the distribution and estimate the mean and error of the distribution. We have seen in the previous section that the systematic error estimates around pericentre are $\Delta V_c\sim3\kms$ and $\Delta q\sim0.04$. We sum these in quadrature with the estimated impact of observational error on the resulting parameter estimates.

We first explore the effects of including an observational error to each observable independently (O1). The current position errors $\sigma_{l,b}=100\mas$ do not produce a significant change to the position of the minimum found using perfect data. Therefore, for all other observations we use a more optimistic position error of $10\mas$. A distance error of $\sim 5$ per cent (O1b) does not alter the estimate of the circular speed significantly but slightly decreases the estimate of $q$, such that it is not consistent with the truth within the quoted error. A proper-motion error of $0.21\masyr$ (which corresponds to a transverse velocity error of $4.7\kms(\kpc\masyr)^{-1}\times0.21\masyr\times10\kpc\approx 10\kms$) applied independently produces a result which is consistent with the truth (O1c). An equivalent line-of-sight velocity error (O1d) shifts the minimum in the circular speed up slightly, and moves the minimum in $q$ down significantly. The effects of errors in these three coordinates are different and probably reflect the observation geometry. We are observing at pericentre at the same azimuthal angle as the centre of the stream. Therefore the line of sight velocity around the centre of the stream (the densest part) will be very small so small errors can have a significant effect. The effect of observational errors on the Galactocentric Cartesian coordinates is also reasonably complex, particularly for the distance error which affects both the observed position and velocity of a star.

When these three errors are combined (O2) the circular speed is shifted
upwards and $q$ is recovered within the errors. Increasing the error in the
proper motion (O3) shifts the circular speed upwards further and also
systematically shifts the minimum in $q$ upwards. An increase in the distance
error (O4) again shifts the circular speed estimate upwards but the recovery
of $q$ is good. O5 shows a combination of errors which recovers the
underlying potential parameters. These errors are perhaps too optimistic and
unlikely to be achieved in the near future. The line of sight velocity error
may be achieved by RAVE, whilst the proper motion error may be achieved by
{\it Gaia}. However it is unlikely that such a small distance error will be
achieved soon from either parallax measurements, or spectroscopic or
photometric distances. However, if one were able to identify standard
candles, such as RR Lyrae stars, in a stream, this distance error is
achievable now \citep{Drake2013}. With a low-mass stream, such as GD-1, we
expect very few RR Lyrae stars, so using this method may only be useful for
more massive streams. Also, we require 500 observations of this quality from
a tidal stream to replicate this test, which makes such a data set even more
improbable. O6 shows an example of large errors. Both the circular speed and
the $q$ estimate are shifted to very large values. However, these results do
not give a good indication of the parameter estimates with these errors as we
are approaching the edge of the allowed range of parameters. We can still see
that large errors degrade the landscape significantly so that no minimum is
found within the appropriate region.

The effect of experimental errors on the position of the minimum is clearly very complex and to use this method with confidence one would need to investigate the systematics on an appropriate simulation before application to data. We see that when large errors are added, the minimum can shift away from the true parameters significantly with very small associated formal errors. The error in the location of these minima does not encompass the true parameter values. In conclusion it seems that current and near-future data could lead to limited use of this method: to use the method with confidence we require accuracy in the observables that may not be attained for many tidal streams for some while. 

\begin{table*}
\caption{Observational errors used to make observations of the stream near pericentre at $t=4.27\Gyr$, along with the estimates of the potential parameters ($V_c, q$) using the stream-fitting algorithm outlined in this paper. $\sigma_{l,b}$ is the observational error in the sky position, $\sigma_s/s$ the fractional distance error, $\sigma_\mu$ the proper motion error, and $\sigma_{||}$ the line-of-sight velocity error. The O\# observations use all 500 stars in the stream sample to determine the potential parameters whilst the OA\# observations bin the observational data in 10 bins on the sky and average before performing the algorithm. The O4M and OA3M observations use a simulation which uses a progenitor cluster which is ten times more massive, but uses the same observational errors as O4 and OA3 respectively. The final two columns give the deviation in standard deviations between the found parameters and the true parameters. At the bottom of the table we show current and future observational errors for comparison. SEGUE data taken from \protect\cite{Pier2003}, \protect\citet{Juric2008}, \protect\citet{Munn2004} and \protect\citet{Yanny2009}. RAVE data taken from \protect\citet{Siebert2011} and \protect\citet{Burnett2011} \citep[RAVE uses position data from 2MASS,][]{Skrutskie2006}. Pan-STARRS data taken from \protect\citet{Kaiser2002} and \protect\citet{Magnier2008}. LSST data taken from \protect\citet{Ivezic2008} for an $r=22$ star. LEGUE (LAMOST Experiment for Galactic Understanding and Exploration) data taken from \protect\citet{LAMOST} (LAMOST uses SDSS data as an input catalogue). {\it Gaia} data taken from \protect\citet{Perryman2001} for $G=20-21$ mag stars. The line-of-sight velocity errors from {\it Gaia} are $\approx 5\kms$ for cool stars. The Gaia-ESO survey \citep{Gilmore2012} aims to record spectra of $\geq10^5$ {\it Gaia} stars which will yield line-of-sight velocity errors of $\approx1\kms$ for cool stars. The fractional error in the distance for Pan-STARRS, LSST and {\it Gaia} are determined from the error in the parallax under a linear approximation.}
\begin{tabular}{lllllllll}
&$\sigma_{l,b}/\mas$&$\sigma_s/s$&$\sigma_{\mu}/\masyr$&$\sigma_{||}/\kms$&$V_c/\kms$&$q$&$\frac{\Delta V_c}{\sigma_{V_c}}$&$\frac{\Delta q}{\sigma_{q}}$\\
\hline\\
O1a			&$100$			&$0.0$			&$0.0$		&$0$		 &$ 220 \pm 3 $&$ 0.92 \pm 0.04 $&$ 0.1 $&$ 0.4 $\\
O1b			&$0$			&$0.04$			&$0.0$		&$0$		 &$ 220 \pm 3 $&$ 0.85 \pm 0.04 $&$ 0.1 $&$ 1.1 $\\
O1c			&$0$			&$0.0$			&$0.21$		&$0$		 &$ 221 \pm 3 $&$ 0.88 \pm 0.04 $&$ 0.2 $&$ 0.4 $\\
O1d			&$0$			&$0.0$			&$0.0$		&$10$		 &$ 222 \pm 3 $&$ 0.83 \pm 0.04 $&$ 0.7 $&$ 1.6 $\\
\\
O2			&$10$			&$0.04$			&$0.21$		&$10$		 &$ 224 \pm 4 $&$ 0.87 \pm 0.04 $&$ 1.2 $&$ 0.7 $\\
O3			&$10$			&$0.04$			&$0.42$		&$10$		 &$ 228 \pm 4 $&$ 0.94 \pm 0.05 $&$ 1.9 $&$ 0.9 $\\
O4			&$10$			&$0.08$			&$0.42$		&$10$		 &$ 231 \pm 5 $&$ 0.92 \pm 0.05 $&$ 2.3 $&$ 0.4 $\\
O5			&$10$			&$0.04$			&$0.21$		&$1$		 &$ 221 \pm 3 $&$ 0.88 \pm 0.04 $&$ 0.4 $&$ 0.5 $\\
O6			&$10$			&$0.16$			&$0.1$		&$1$		 &$ 261 \pm 8 $&$ 1.05 \pm 0.08 $&$ 5.3 $&$ 1.9 $\\
\\
OA1			&$10$			&$0.04$			&$0.21$		&$10$		 &$ 224 \pm 4 $&$ 0.9 \pm 0.04 $&$ 1.1 $&$ 0.1 $\\
OA2			&$10$			&$0.08$			&$0.42$		&$10$		 &$ 224 \pm 4 $&$ 0.9 \pm 0.05 $&$ 0.9 $&$ 0.0 $\\
OA3			&$10$			&$0.2$			&$1.2$		&$10$		 &$ 223 \pm 10 $&$ 0.91 \pm 0.09 $&$ 0.3 $&$ 0.1 $\\
\\
O4M			&$10$			&$0.08$			&$0.42$		&$10$		 &$ 225 \pm 4 $&$ 0.91 \pm 0.05 $&$ 1.5 $&$ 0.2 $\\
OA3M		&$10$			&$0.2$			&$1.2$		&$10$		 &$ 223 \pm 6 $&$ 0.9 \pm 0.07 $&$ 0.5 $&$ 0.0 $\\
\hline\\

SEGUE		&$50-100$		&$0.15-0.2$			&$3$			&$10$		&$$&$$\\
RAVE		&$\la100$		&$0.133$			&$\la 8$		&$<2$		&$$&$$\\
Pan-STARRS	&$\sim10$		&$15(s/\kpc)$		&$1.2$			&$-$		&$$&$$\\
LEGUE		&$\la100$		&$-$				&$-$			&$1-7$		&$$&$$\\
LSST		&$15$			&$0.8(s/\kpc)$		&$0.3$			&$-$		&$$&$$\\
{\it Gaia}		&$0.14-0.44$	&$<0.5(s/\kpc)$	&$0.12-0.38$	&$\sim1-5$	&$$&$$\\
\hline
\end{tabular}
\label{ObsErrors}
\end{table*}

\section{Data averaging in observable space}
Despite the results of the previous section, the method could still prove relevant if slightly adjusted. The motion of the stars in a stream in real space is coherent, and the above algorithm does not utilise this fact. This additional information may be exploited by first binning in real space. The stream is essentially a one dimensional structure regardless of whether it delineates an orbit or not. Therefore binning along the stream is valid even when the stream does not delineate the orbit. We form bins along the stream and estimate the observables at the centre of the bin by the average of the observables of the stars in the bin. One of the advantages of the above routine is that the linear regression in angle-space and frequency-space beats down the random errors in the estimated angles and frequencies. Averaging the data in observable space on the sky provides an alternative method for reducing the random errors, which for large observational errors is probably the preferable technique. Indeed, when analysing the GD-1 stream \citet{Koposov2010} mitigate the problem of large errors by binning and averaging the data along the stream in observable space to obtain a handful of much more accurate data points -- the proper-motion errors for instance can be reduced by a factor of five. 

%The relationship between observational errors and angle/frequency errors is complex and non-linear. For instance, for the observation explored above, small errors in the line-of-sight velocity introduce larger errors in the resulting potential parameters than the equivalent proper motion errors. These effects are obviously magnified for larger observational errors. Therefore smaller input observational errors are more desirable as the expected structure is more likely to be recovered. Data with very large observational errors are less likely to produce clean structures in angle- and frequency-space.

When large observational errors are added, particles can be scattered to very high actions. \citet{Sanders2012} showed that the systematic errors in the angle-action variables increases with the actions. Therefore larger observational errors will also introduce larger systematic errors. In the extreme case, observational errors can make orbits unbound\footnote{A logarithmic potential does not tend to an asymptotic limit at large $R$ or $z$ so technically every orbit in a logarithmic potential is bound. However, in practice, orbits which reach very large $R$ and $z$ take an extremely long time to integrate. Therefore, we consider all orbits which stray more than $200\kpc$ from the Galactic centre to be `unbound'.}. These orbits have infinite actions and must be ignored when performing the linear regression in angle- and frequency-space. Averaging data in observable space should remove these issues as we will have a data set which spans a smaller range in actions.

Unfortunately, averaging in observable space results in a smaller number of data points with which to perform the linear regression. In the above examples we have used $500$ data points which provides a very tight linear regression. However, even just a handful of data points should provide a good estimate of the gradient.

Here we explore the results of observational data averaging for three observations listed in Table~\ref{ObsErrors}. We bin the `observed' data in each observational coordinate using ten equally populated bins in Galactic longitude, $l$. The bins are sufficiently small for the stream to be approximately linear in each subspace for each bin so averaging within a bin works well. This procedure reduces the error in the input coordinates by a factor of $\sim 1/\sqrt{50} \approx 0.14$. After transforming back to the Galactocentric Cartesian coordinates, we perform the algorithm on the averaged data to find the best combination of potential parameters. We then repeat the scattering and binning process $100$ times to estimate the error. In Fig.~\ref{Observed} we plot the binned data for the observation OA3.

We can see from Table~\ref{ObsErrors} that averaging in observable space produces superior results. First the systematic shifts in the minima are removed, and we recover an average for the parameters which is close to the truth, particularly for $q$. The average position of the minimum does not depend as sensitively on the magnitude of the observational errors (compare OA1 and OA2). If we compare O4 and OA2 the observation errors are the same, but the recovery of the parameter estimates with the two methods are different: the circular speed is better recovered by averaging in observable space. This is achieved without an increase in the quoted error of the parameter estimates. Second we can weather much larger observational errors. The errors in observation OA3 seem achievable with future data and we correctly recover the potential parameters using the averaging method. For all three observations the estimated errors encompass the true potential parameters.

In conclusion we can achieve the most accurate results by first averaging in
observable space and then performing the algorithm on the reduced data set.
This is the correct method to use as we remove unphysical measurements (those
with infinite actions) and hence we only spend time calculating the angles
and frequencies for reliable phase-space points. Similarly we need only
calculate the angles and frequencies -- a time-consuming process -- for a
handful of points. Scattering and binning is,
in comparison, very fast. These principles apply to any process trying to
calculate actions, or any other non-trivial variable, from noisy data:
binning the data in the space of observables removes outliers and results in
fewer, more reliable points to run through the complicated calculation.

Finally we briefly investigate how the recovery of the potential parameters
depends upon the initial mass of the cluster. In Paper I we showed that the
stream-orbit misalignment is independent of the cluster mass. The effect of
increasing the cluster mass is to scale the angle and frequency
distributions without altering the morphology. We take the
$M_c=2\times10^5M_\odot$ simulation from Paper I, which was evolved on the
same orbit for the same time as the above simulation. Due to the higher mass
the resulting stream is longer, spanning $\sim190^\circ$ in Galactic
longitude, so we expect that the estimates of the parameters will be
superior as the stream probes a larger range in both angles and frequencies.
In Table~\ref{ObsErrors} we show the estimates of the parameters from this
higher-mass simulation using the observational errors from O4 and OA3 with
the observable-space binning (these are labelled in the table as O4M and
OA3M). For O4M the increased stream length has weakened the systematic shift
of the circular velocity found in O4, producing a mean value which is closer
to the truth. Despite the increased width of the stream the error in the
parameters is small. For OA3M the recovery of parameters is similar to the
OA3 observation, but the error in these estimates has been reduced without
excluding the truth from the error ellipse. These two experiments show that longer streams, which have a larger spread in angle space, produce superior parameter estimates.

\section{Conclusions}

Tidal streams are very attractive structures for probing the Galactic potential on large scales, but it is essential that appropriate algorithms are developed for their study. In Paper I we showed that orbit-fitting is inappropriate for many streams in the Milky Way, and can lead to order one errors in estimation of parameters of the Galactic potential.

Motivated by the need for an improvement over orbit-fitting, we have presented an algorithm for using tidal-stream data to constrain the potential of the Galaxy without assuming that the stream delineates an orbit. Instead it identifies the true potential as the one that yields corresponding patterns in angle and frequency space. The algorithm was tested by evolving an N-body simulation of a King cluster in a two-parameter logarithmic potential until a stream is formed. The degree of correlation between the angle and frequency structure was maximised with respect to trial potentials. The correct parameters were recovered within the estimated systematic errors of the method.

As tidal streams are very distant structures we expect large errors in the observables. Therefore, it is imperative that any stream-fitting method is shown to function for large observational errors. We have shown that the observational errors in the distances and proper motions for individual stars in tidal streams are currently too large to use the above technique with any confidence. However, if the data are first binned and averaged in observable space on the sky, we can recover the correct potential parameters even for large observational errors. The current observational errors may be small enough for close streams such as GD-1 and it seems promising that in the era of ${\it Gaia}$ the data for more streams will be sufficiently accurate to use this algorithm.

We have shown that longer streams produce superior potential parameter estimates, so there is hope for using higher-mass streams to produce better constraints on the potential. We have seen that observing a stream at different times results in different constraints on the potential. Therefore there is much to be gained by using several streams simultaneously to constrain the potential. Hopefully this would also remove local minima and make a global minimum more apparent. Similarly the approach taken here uses a constant prior for the potential parameters. In reality, a more informative prior could be used, that would rule out regions of the parameter space which are populated by local minima.

We have only investigated how well a very simple two-parameter potential may be constrained. In reality, models of the Galactic potential have many more parameters. In this case we expect many more degeneracies in parameter space and a much fuller Monte Carlo search of this higher dimensional space would be required. From the test results here we have seen that we can constrain the circular speed of the potential more effectively than the shape. It would be interesting to explore with a more complex potential which parameters/features of the potential we are able to constrain more accurately than others. For instance, when applying an orbit-fitting algorithm to the GD-1 stream \citet{Koposov2010} found that the mass of the disc in their Galaxy model could be more accurately constrained than the shape of the halo.

Finally, this algorithm, unlike an orbit-fitting algorithm, requires full 6D data for the stream. The Galaxy is being increasingly mapped in 6D and we have 6D data for the GD-1 stream \citep{Koposov2010}. However it may be some time before we have accurate 6D data for many distant streams. Therefore, to make the presented algorithm more usable in the near future it would be interesting to investigate what alterations we can make to deal with reduced phase-space information. We intend to adapt the above approach to produce a Bayesian model of the stream structure in angle-action space, which will be useful in this respect.

\section*{Acknowledgements}
JS acknowledges the support of the Science and Technology Facilities Council and we thank Paul McMillan for useful discussions. We also thank the anonymous referee for their useful suggestions which have improved the paper.
{\footnotesize{
\bibliographystyle{mn2e-2}
\bibliography{StreamFit}
}}
\appendix
\section{Locally Fitting St\"ackel potentials}\label{StackelFit}
\citet{Sanders2012} presented a method for estimating the angle-action variables in a general axisymmetric potential given a 6D phase-space point. It is this method which we use in this paper to determine the angle- and frequency-space structure of the stream. The method proceeded by first finding the region the point would explore as it proceeded along its orbit in the potential. The given potential is then approximated by a St\"ackel potential inside this region. Once an appropriate St\"ackel potential is found, the true actions, angles and frequencies may be estimated by those calculated in the St\"ackel potential. St\"ackel potentials are the most general class of potentials for which angle-action variables can be calculated analytically. In this appendix we present the relevant equations for determining actions, angles and frequencies in a St\"ackel potential. For more information on the algorithm readers should consult \citet{Sanders2012}.

In a St\"ackel potential the Hamilton-Jacobi equations separate which allows us to write down an expression for the third isolating integral, $I_3$ \citep{deZeeuw1985}. Axisymmetric St\"ackel potentials are associated with prolate spheroidal coordinate systems defined by two constants $(a,c)$. These coordinates are related to cylindrical polar coordinates $(R,\phi,z)$ by
\begin{equation}
\frac{R^2}{\tau-a^2}+\frac{z^2}{\tau-c^2} = 1,
\end{equation}
where $\lambda$ and $\nu$ are the roots of $\tau$ such that $c^2\leq\nu\leq a^2\leq\lambda$. Surfaces of constant $\lambda$ are prolate spheroids and surfaces of constant $\nu$ are two-sheeted hyperboloids of revolution which intersect the spheroids orthogonally. An axisymmetric potential is of St\"ackel form if
\begin{equation}
\Phi_S = -\frac{f(\lambda)-f(\nu)}{\lambda-\nu}.
\label{StackDef}
\end{equation}
$\Phi_S$ is fully defined by a single function $f(\tau)$. A single function may be used as $\lambda$ and $\nu$ take different ranges of values except at $\lambda=a^2, \nu=a^2$, where we require $f$ to be continuous so the potential remains finite.
The momentum $p_\tau$ is solely a function of its conjugate variables, $\tau$, and the three isolating integrals in a St\"ackel potential, (the energy $E$, the $z$-component of the angular momentum $L_z$ and $I_3$):
\begin{equation}
2(\tau-a^2)p_\tau^2 = E-\frac{L_z^2}{2(\tau-a^2)}-\frac{I_3}{\tau-c^2}+\frac{f(\tau)}{\tau-c^2}.
\label{TauEqOfMotion}
\end{equation}
We define the action variables, $J_R$ and $J_z$, as
\begin{equation}
J_R = \frac{1}{2\pi}\oint p_\lambda\mathrm{d}\lambda,\quad J_z = \frac{1}{2\pi}\oint p_\nu\mathrm{d}\nu
\label{actionDef}
\end{equation}
where the integration is over all values of $\tau$ for which $p_\tau^2\geq0$. The third action, $J_\phi$, is simply $L_z$.
The corresponding angle coordinates, $\boldsymbol{\theta} = (\theta_R,\theta_\phi,\theta_z)$, are calculated by the introduction of the generating function, $S(\lambda,\phi,\nu,J_R,L_z,J_z)$ for the canonical transformation from $(\lambda,\phi,\nu,p_\lambda,p_\nu,L_z)$ to $(\theta_R,\theta_\phi,\theta_z,J_R,L_z,J_z)$. The angles are found by differentiating the generating function with respect to the respective action such that
\begin{equation}
\theta_R = \frac{\partial S}{\partial J_R},\quad \theta_z = \frac{\partial S}{\partial J_z},\quad\theta_\phi=\frac{\partial S}{\partial L_z}.
\end{equation}
The frequencies, $\boldsymbol{\Omega}$, are related to the derivatives of the actions with respect to the integrals of the motion by
\begin{equation}
\begin{split}
\Omega_R &= \frac{1}{\Gamma}\frac{\partial J_z}{\partial I_3},\\
\Omega_\phi &= \frac{1}{\Gamma}\Big(\frac{\partial J_z}{\partial L_z}\frac{\partial J_R}{\partial I_3}-\frac{\partial J_z}{\partial I_3}\frac{\partial J_R}{\partial L_z}\Big),\\
\Omega_z &= -\frac{1}{\Gamma}\frac{\partial J_R}{\partial I_3},
\end{split}
\end{equation}
where 
\begin{equation}
\Gamma = \Big(\frac{\partial J_R}{\partial E}\frac{\partial J_z}{\partial I_3}-\frac{\partial J_R}{\partial I_3}\frac{\partial J_z}{\partial E}\Big).
\end{equation}
Expressions for all these derivatives are given in \citet{Sanders2012}. 
% The accuracy of estimating angle-action variables in this way is discussed in \cite{Sanders2012} but the accuracy of estimating the frequencies is not discussed. %Here we briefly discuss how accurately the frequencies can be recovered. 
% To use the algorithm presented in this paper the frequencies along a section of an orbit, approximately the size of the stream, should be sufficiently constant and there should be no correlation between the frequency components. We show that we can determine the frequencies for the orbit used in the tests to a sufficient accuracy in Fig.~\ref{ClusterEndStateAA} and give an estimate of the errors below.

\subsection{Estimating $I_3$}
In the algorithm presented in \citet{Sanders2012} $I_3$ was estimated by averaging three estimates along the orbit. $I_3$ is an integral of the motion in a St\"ackel potential so in that paper it was desirable to avoid the estimates of $I_3$ depending on the initial position on the orbit. This produced much smaller errors in the estimated actions so was a favourable scheme for the task presented. However, problems arise around the turning points of the orbit with a poor choice of $I_3$.  A small error in $I_3$ can move the boundary of the orbit away from the true boundary, so points which are close to the boundary appear further away than they should be. As $I_3$ is a separation constant it can be estimated in two different ways, using $\dot{\lambda}$ or $\dot{\nu}$:
\begin{equation}
\begin{split}
I_3 &\approx (\lambda-c^2)\Bigg(E-\frac{L_z^2}{2(\lambda-a^2)}+\frac{f(\lambda)}{\lambda-c^2}-\frac{\dot{\lambda}^2(\lambda-\nu)^2}{8(\lambda-a^2)(\lambda-c^2)^2}\Bigg),\\
&\approx (\nu-c^2)\Bigg(E-\frac{L_z^2}{2(\nu-a^2)}+\frac{f(\nu)}{\nu-c^2}-\frac{\dot{\nu}^2(\nu-\lambda)^2}{8(\nu-a^2)(\nu-c^2)^2}\Bigg).
\end{split}
\end{equation}
In a St\"ackel potential these two estimates are identical but in a general potential we expect a discrepancy. Using only the $\dot{\lambda}$ estimate leads to errors in $\theta_z$ near the turning points and similarly using only the $\dot{\nu}$ estimate leads to errors in $\theta_R$ near the turning points.

When analysing the stream data we are primarily interested in the angles and the behaviour of the stream near apsis. Also, we are only interested in relative differences between stars in the stream. Therefore, we use a modified scheme to that presented in \citet{Sanders2012}. We choose to only estimate $I_3$ at the given phase-space point. We also calculate two different estimates for $I_3$ using the $\dot{\lambda}$ and $\dot{\nu}$. The equation of motion for $\lambda$ uses the $\dot{\lambda}$ estimate and similarly for $\nu$. In this way we remove noise around the turning points and the stream structure is recovered much more cleanly.

Several other small alterations have been made to the algorithm presented in \citet{Sanders2012}, which improve the performance for the present task. We have increased the number of $\alpha$-estimates per particle, and increased the size of the potential-fitting region and selected the weight function $\Lambda(\lambda) = 3\lambda^{-4}(\lambda^{-3}_{+}-\lambda^{-3}_{-})^{-1}$ to reduce noise in the frequencies.

\subsection{Impact of errors on suggested algorithm}\label{ErrorsAppendix}
The St\"ackel-fitting algorithm used to estimate the angles and frequencies introduces errors, which systematically depend on the orbital phase. Using the same method to estimate the errors as that presented in \cite{Sanders2012}, we find that the standard errors in the angles for the orbit used in Section~\ref{Test} are $(\sigma_{\theta_R},\sigma_{\theta_\phi}, \sigma_{\theta_z}) = (9.4,5.7,5.4)\times 10^{-3}\rad$. For a single orbit the frequencies should be constant at all points along the orbit. We estimate the error in the calculated frequencies as the standard deviation of their estimates around the orbit which yields $(\sigma_{\Omega_R},\sigma_{\Omega_\phi}, \sigma_{\Omega_z}) = (3.7,4.0,2.6)\times 10^{-3}\Gyr^{-1}$. These errors are much smaller than the size of the distributions shown in Fig.~\ref{ClusterEndStateAA}. The errors in the frequency and angle estimation depend upon the phase of the particle, with the largest error occurring at pericentre.

However, these absolute errors do not give a good indication of the accuracy of the algorithm presented in this paper. As we are measuring the gradients of the frequency distribution and the angle distribution, we are instead concerned with the errors in the differences of the frequencies/angles of the particles in the stream, which are subject to more subtle effects. For simplicity we limit the discussion to the errors in the frequencies but similar effects are found in the angles. Let us consider two particles in the stream: one in the cluster with frequency $\boldsymbol{\Omega}_0=(\Omega_{R0},\Omega_{\phi 0},\Omega_{z0})$ and another particle, which has just been stripped from the progenitor with frequency $\boldsymbol{\Omega}=(\Omega_{R0}+\Delta\Omega_R,\Omega_{\phi 0}+\Delta\Omega_\phi,\Omega_{z0}+\Delta\Omega_z)$. The St\"ackel-fitting algorithm produces deviations in the frequencies, given by $\boldsymbol{\sigma}(t,\boldsymbol{\Omega})$, which depend on the orbital phase so fluctuate in time at a rate determined by the frequencies. 

\begin{figure}
$$\includegraphics{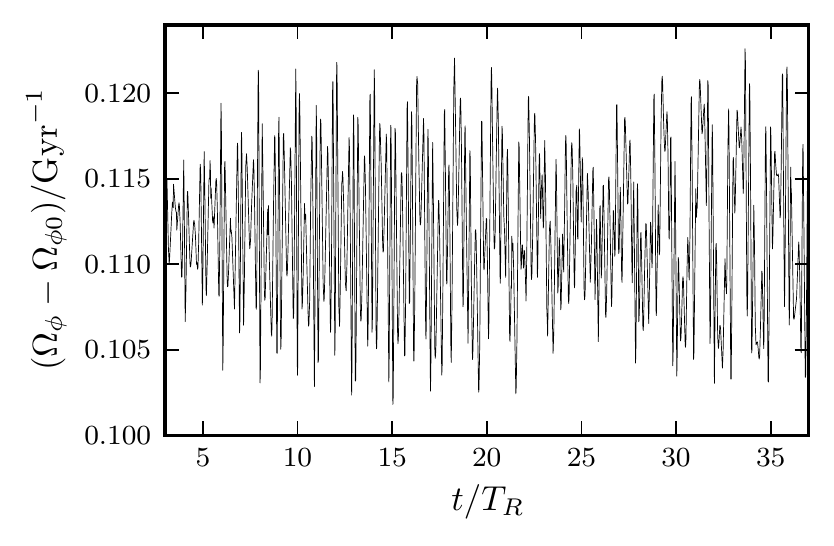}$$
\caption{The azimuthal frequency difference between the progenitor and a single particle taken from the simulation, and integrated for a long time. For all shown times the particle is moving freely in the external Galactic potential. The error in the frequency difference exhibits beats at a frequency of $\sim 0.3\Gyr^{-1}$ characteristic of two particles oscillating at close frequencies drifting in and out of phase. The units give the times of pericentric passage of the progenitor.}
\label{FreqDiff}
\end{figure}

At a given time, the difference in the estimated frequencies is given by
\begin{equation}
\begin{split}
\delta\Omega_i &= [\Omega_i+\sigma_i(t,\boldsymbol{\Omega})]-[\Omega_{0i}+\sigma_i(t,\boldsymbol{\Omega}_0)]\\
&\approx\Delta\Omega_i+2g_i(\Delta\boldsymbol{\Omega},t),
\end{split}
\end{equation}
where $g_i(\Delta\boldsymbol{\Omega},t)$ is an oscillating function of the frequency difference, and we have dropped the fast-oscillating part.
Therefore, there is an error in the difference which oscillates in time at a frequency related to the beat frequency. In Fig.~\ref{FreqDiff} we plot the azimuthal frequency difference, $\Delta\Omega_\phi$, as a function of time for a single particle from the simulation integrated for a long time. The particle was stripped at the third pericentric passage and then integrated independent of the simulation for many more periods. We observe that the fluctuations in the frequency introduced by the algorithm are producing beats in $\Delta\Omega_\phi$ with a frequency of $\sim 0.3\Gyr^{-1}$. We expect that at early times $t\ll 5\Gyr$ the error in the difference will be small as the systematic errors in the progenitor and particle frequency-estimates are in phase and so nearly cancel. At large times when $t\approx 5\Gyr$ the error fluctuations in the particle and progenitor frequencies have drifted out of phase, so the error in the difference is large. The error in $\Delta\Omega_\phi$ reaches a maximum of $\sim 10\percent$. A consequence of this systematic is that, at early times, the particles with the largest frequency difference will introduce the largest errors, as they were the first to leave the cluster. This explains the broadening of the extremes of the frequency distributions in Fig.~\ref{ClusterEndStateAA}. A similar effect is observed in the angles shown in Fig.~\ref{AngleswithTimes}. The largest error in the angles occurs at pericentre \citep{Sanders2012}. Around each pericentric passage the St\"ackel-fitting algorithm produces a small blip in the angle difference. This blip increases in magnitude with time as the systematic errors introduced by the algorithm shift out of phase with each other.

The discussion has been limited to the errors introduced when finding the gradient using a single particle and the progenitor. It is more difficult to assess how the linear regression of the angles and frequencies of many particles in the stream is affected, but it is clear that it is systematic and not random, as the errors of neighbouring stream particles are correlated. The errors depend upon the initial cluster conditions which govern $\Delta\boldsymbol{\Omega}$, the time since the cluster started being stripped, and the phase at which the stream is observed. We also expect this source of error to decrease with the mass of the progenitor. Larger progenitor masses produce larger frequency distributions (see Paper I), so the errors introduced by the St\"ackel-fitting algorithm become less significant.

\bsp

\label{lastpage}

\end{document}